\documentstyle[12pt,epsfig]{article}                                                
                                                
\newlength{\dinwidth}                                                
\newlength{\dinmargin}                                                
\def\lapproxeq{\lower .7ex\hbox{$\;\stackrel{\textstyle                                                
<}{\sim}\;$}}                                                
\def\gapproxeq{\lower .7ex\hbox{$\;\stackrel{\textstyle                                                
>}{\sim}\;$}}                                               
\def\funp{{I\!\! P}}                                         
\def\be{\begin{equation}}                                                
\def\ee{\end{equation}}                                                
\def\bea{\begin{eqnarray}}                                                
\def\eea{\end{eqnarray}}                                                
                                                
\makeatletter                                                
\def\fmslash{\@ifnextchar[{\fmsl@sh}{\fmsl@sh[0mu]}}                                                
\def\fmsl@sh[#1]#2{%
\mathchoice                                                
{\@fmsl@sh\displaystyle{#1}{#2}}%
{\@fmsl@sh\textstyle{#1}{#2}}%
{\@fmsl@sh\scriptstyle{#1}{#2}}%
{\@fmsl@sh\scriptscriptstyle{#1}{#2}}}                                                
\def\@fmsl@sh#1#2#3{\m@th\ooalign{$\hfil#1\mkern#2/\hfil$\crcr$#1                                                
#3$}}                                                
\makeatother                                                
                                                
\begin{document}                                                
\titlepage                                                
\begin{flushright}                                                
DTP/98/20 \\                                                
May 1998 \\                                                
\end{flushright}                                                
                                                
\begin{center}                                                
\vspace*{2cm}                                                
{\Large \bf The description of $F_2$ at low $Q^2$}                                                
\\                                                
\vspace*{1cm}                                                
A.D.\ Martin$^1$, M.G.\ Ryskin$^{1,2}$ and A.M.\ Stasto$^{1,3}$                                                
\end{center}                                                
                                                
\vspace*{0.5cm}                                                
                                                
\begin{tabbing}                                                
$^1$xxxxx\= \kill                                                
                                                 
\indent $^1$ \> Department of Physics, University of Durham,                                                
Durham, DH1 3LE, UK. \\                                                
                                                
\indent $^2$ \> Petersburg Nuclear Physics Institute, 188350,                                                
Gatchina, St. Petersburg, Russia. \\                                                
                                                
\indent $^3$ \> H.\ Niewodniczanski Institute of Nuclear Physics,                                                
31-342 Krakow, Poland.  \\                                                
\end{tabbing}                                                
                                                
\vspace*{2cm}                                                
                                                
\begin{abstract}                                                
We analyse the data for the proton structure function $F_2$ over the entire $Q^2$                                     
domain, including especially low $Q^2$, in terms of perturbative and                                                
non-perturbative QCD contributions.  The small distance                                                
configurations are given by perturbative QCD, while the large                                                
distance contributions are given by the vector dominance model                                                
and, for the higher mass $q \overline{q}$ states, by the additive                                                
quark approach.  The interference between states of different $q\bar{q}$ mass (in the  
perturbative contribution) is found to play a crucial role in obtaining an excellent  
description of the data throughout the whole $Q^2$ region, including  
photoproduction.  
\end{abstract}                                                
                                                
\newpage                                                
\noindent{\large \bf 1.~Introduction}                                                
                                                
There now exist high precision deep inelastic $ep$ scattering                                                
data \cite{H1,ZEUS} covering both the low $Q^2$ and high $Q^2$                                                
domains, as well as measurements of the photoproduction cross                                                
section.  The interesting structure of these measurements, in                                                
particular the change in the behaviour of the cross section with $Q^2$ at $Q^2 \sim                                     
0.2 {\rm GeV}^2$, highlight the importance of obtaining a theoretical QCD                                                
description which smoothly links the non-perturbative and perturbative domains.                                                
                                                
In any QCD description of a $\gamma^* p$ collision, the first                                                
step is the conversion of the initial photon into a $q                                                
\overline{q}$ pair, which is then followed by the interaction of                                                
the pair with the target proton.  Let $\sigma (s, Q^2)$ be the                                                
total cross section for the process $\gamma^* p \rightarrow X$                                                
where $Q^2$ is the virtuality of the photon and $\sqrt{s}$ is the                                                
$\gamma^* p$ centre-of-mass energy.  It is related to th 
e forward $\gamma^*                                                
p$ elastic amplitude $A$ by the optical theorem, ${\rm Im} \; A                                                
\: = \: s \sigma$.  We may write a double dispersion relation                                                
\cite{DDR} for $A$ and obtain for fixed $s$                                                 
\be                                                
\label{eq:a1}                                                
\sigma (s, Q^2) \; = \; \sum_q \: \int \: \frac{d M^2}{M^2 + Q^2} \: \int                                                
\: \frac{d M^{\prime 2}}{M^{\prime 2} + Q^2} \; \rho (s, M^2,                                                
M^{\prime 2}) \: \frac{1}{s} \; {\rm Im} \; A_{q \overline{q} +                                                
p}                                                
(s, M^2, M^{\prime 2})                                                
\ee                                                
where $M$ and $M^\prime$ are the invariant masses of the incoming and outgoing                                             
$q\bar{q}$ pair.  The relation is shown schematically in Fig.~1.  If we assume that                                                
forward $q \overline{q} + p$ scattering does not change the                                                
momentum of the quarks\footnote{In a more detailed treatment this                                                
assumption is no longer valid, see (\ref{eq:d17}) and                                                
(\ref{eq:a19}) below, and the discussion in section 4.} then                                                
$A_{q \overline{q} + p}$ is                                                
proportional to $\delta (M^2 - M^{\prime 2})$, and (\ref{eq:a1})                                                
becomes                                                
\be                                                
\label{eq:a2}                                                
\sigma (s, Q^2) \; = \; \sum_q \: \int_0^\infty \: \frac{d M^2}{(M^2 +                                                
Q^2)^2} \; \rho (s, M^2) \: \sigma_{q \overline{q} + p} (s, M^2)                                                
\ee                                                
where the spectral function $\rho (s, M^2)$ is the density of $q                                                
\overline{q}$ states.                                                
                                                
Following Badelek and Kwiecinski \cite{BK} we may divide the                                                
integral into                                                
two parts\footnote{Although Badelek and Kwiecinski base their fit to the data on                            
eq.~(\ref{eq:a2}), they also discuss                                                
the more general case in which $M \neq M^\prime$ contributions                                                
may be included in the spectral function $\rho$.}, the region $M^2 < Q_0^2$                                                 
described by the vector meson                                                
dominance model (VDM) and the region $M^2 > Q_0^2$ described by                                                
perturbative QCD.  Suppose that we assume $\rho \sigma_{q                                                
\overline{q} + p}$ is a constant independent of $M^2$ (which                                                
should be true modulo logarithmic QCD corrections) then the                                                
perturbative component of the integral is                                                
\be                                                
\label{eq:a3}                                                
\int_{Q_0^2}^{\infty} \frac{d M^2}{(M^2 + Q^2)^2} \: \rho \sigma                                                
\; = \; \int_0^\infty \frac{d M^2}{(M^2 + Q^2 + Q_0^2)^2} \: \rho                                                
\sigma \; = \; \sigma (s, Q^2 + Q_0^2).                                                
\ee                                                
Thus (\ref{eq:a2}) becomes                                                
\be                                                
\label{eq:a4}                                                
\sigma (s, Q^2) \; = \; \sigma ({\rm VDM}) + \sigma^{{\rm QCD}}  \:  (s, Q^2 +                                                
Q_0^2)                                                
\ee                                                
where the QCD superscript indicates that the last contribution is to be calculated                                                 
entirely from perturbative QCD.                                                
                                                
We may use                                                
\be                                                
\label{eq:a5}                                                
\sigma (s, Q^2) \; = \; \frac{4 \pi^2 \alpha}{Q^2} \: F_2 (x,                                                
Q^2)                                                
\ee                                                
where $x = Q^2 / (s + Q^2 - M^2)$ to rewrite (\ref{eq:a4}) as                                                
\be                                                
\label{eq:a6}                                                
F_2 (x, Q^2) \; = \; F_2 ({\rm VDM}) + \frac{Q^2}{Q^2 + Q_0^2} \:                                                
F_2^{{\rm QCD}}  \: (\overline{x}, Q^2 + Q_0^2)                                                
\ee                                                
where $\overline{x} = (Q^2 + Q_0^2) / (s + Q^2 + Q_0^2 - M^2)$.                                                 
The vector meson dominance term has the form\footnote{Strictly                                                
speaking (\ref{eq:a7}) is the formula $F_T$.  The small                                                
longitudinal component will be discussed later.}                                                
\be                                                
\label{eq:a7}                                                
F_2 ({\rm VDM}) \; = \; \frac{Q^2}{4 \pi} \; \sum_V \:                                                
\frac{M_V^4 \: \sigma_V \: (s)}{\gamma_V^2 (Q^2 + M_V^2)^2}                                                
\ee                                                
where $M_V$ is the mass of vector meson $V$ and where the sum is over the                                                 
vector mesons which fall in the region                                                
$M_V^2 < Q_0^2$.  The vector meson-proton cross sections                                                
$\sigma_V (s)$ can be determined from the $\pi p$ and $K p$ total                                                
cross sections using the additive quark model and $\gamma_V^2$                                                
from the leptonic width of the vector meson $V$.  The last term                                                
in (\ref{eq:a6}) can be determined from perturbative QCD using                                                
the known parton distributions.  This approach was first proposed                                                
by Badelek and Kwiecinski (BK) \cite{BK}.  We see that the BK                                                
model, (\ref{eq:a4}) and (\ref{eq:a6}), makes a parameter free                                                
prediction of $F_2 (x, Q^2)$ which is expected to be valid, for                                                
$s \gg Q^2$, for all $Q^2$ including very low $Q^2$.  The BK predictions give an                            
excellent description of the $F_2$ data for $Q^2 \gapproxeq 1$~GeV$^2$, but                            
overshoot the new measurements of $F_2$ for smaller values of $Q^2$.  This                            
deficiency of the model was removed in a fit to the $F_2$ data performed by the H1                            
collaboration \cite{H1}, but at the expense of using an unreasonably low value for                            
$Q_0^2 = 0.45 {\rm GeV}^2$ and of introducing an ad hoc factor of 0.77 to suppress                            
the VDM term.                           
                                                
The Badelek-Kwiecinski idea to separate perturbative                                                
and non-perturbative contributions is very attractive.  To                                                
exploit it further we must achieve a better separation between                                                
the short and long distance contributions.  To do this we take a                                                
two-dimensional integral over the longitudinal and transverse                                                
momentum components of the quark, rather than simply over the                                                
mass $M$ of the $q \overline{q}$ pair.                                                
                                                
The contribution coming from the small mass region is pure VDM                                                
and is given by (\ref{eq:a7}).  However, the behaviour of the                                                
cross section at large $M^2$ is a more delicate question.  The                                                
part which comes from large $k_T$ of the quark can be calculated                                                
by perturbative QCD in terms of the known parton distributions,                                                
whereas for small $k_T$ we will use the additive quark model and                                                
the impulse approximation.  That is only one quark interacts with                                                
the target and the quark-proton cross section is well                                                
approximated by one third of the proton-proton cross section.                                         
                                                
At this point it is interesting to note some recent excellent parametric fits of the data  
for                                     
$F_2$, or rather for $\sigma (\gamma^* p)$.  The first is                                          
based on (\ref{eq:a2}) and the generalised VDM \cite{SS}.  To be more precise it is                                          
based entirely on a parametrization of                                                
the vector meson + proton cross section and does not take                                                
advantage of our present knowledge of perturbative QCD.  As a                                                
consequence some anomalies appear.  For instance the                                                
photoproduction cross section becomes negative for $\sqrt{s} < 6                                                
{\rm GeV}$ (or $\sigma (Vp) < 0$ for $M_V > 0.26 \sqrt{s})$.                                                 
Second the model has anomalously large values of $R = \sigma_L /                                                
\sigma_T$ (where $F_L$ is obtained by including a factor $\xi Q^2/M_V^2$ on the                                     
right-hand-side of formula (\ref{eq:a7}) for $F_T$).  In the well-known deep inelastic                                     
region the model predicts $R > 1$ for $Q^2 > 35 {\rm GeV}^2$ and $x > 0.01$ (and                                                
even $R > 4$ for $x > 0.1$) whereas the data indicate that $R                                                
\simeq 0.2 - 0.3$.  This effect probably reflects, as the authors                                                
note, the omission of allowing $\xi$ to depend on $Q^2$, see                                                
(\ref{eq:a22}) below.  Rather their model has $\xi = 0.171$ for all                                                
$Q^2$.  
 
An earlier approach based on the generalised VDM can be found in  
ref.~\cite{SHAW}.  In addition to the VDM contributions, this work contains a  
contribution at small $x$ coming from \lq\lq heavy" long-lived fluctuations of the  
incoming photon, which are parametrized in terms of a \lq\lq hard" Pomeron whose  
intercept is found to be $\alpha_{P^\prime} = 1.289$. 
                                         
Another fit \cite{AL} of the $F_2$ data is based on the Regge motivated ALLM                                          
parametrization \cite{ALLM}.  The description, with 23 parameters, describes the                                          
data well and may be used to interpolate the measurements.  On the other hand the                                          
physical basis of the parametrization is not clear.  For example a variable $x_\funp$ is                                          
defined by                                         
\begin{equation}                                         
\frac{1}{x_\funp} \; = \; 1 \: + \: \frac{W^2 - M^2}{Q^2 + M_\funp^2}                                         
\label{eq:z2a}                                         
\end{equation}                                         
where $W = \sqrt{s}$ is the $\gamma^* p$ centre-of-mass energy, $M$ is the proton                                     
mass and                                          
$M_\funp$ reflects the energy scale of Pomeron exchange.  This latter scale turns out                                          
to be extremely large, $M_\funp^2 = 49.5$~GeV$^2$, much larger than any hadron or                                          
glueball mass.  Secondly the intercept, $\alpha_R (0)$, of the secondary trajectory                                      
decreases with $Q^2$, which is contrary to Regge theory (where $\alpha_R$ is  
independent of $Q^2$).                                         
                                         
The description of the $F_2$ or $\sigma (\gamma^* p)$ data presented in this paper is                                     
quite different.  We use a physically                                          
motivated approach with very few free parameters, and we clearly separate the                                          
contributions to $F_2$ coming from the large (small quark $k_T$) and small (large                                          
$k_T$) distances. A recent study with a similar philosophy to ours can be found in                               
ref.~\cite{GLM}.  They achieve a qualitative description of the experimental data over                             
a wide range of photon virtualities $(Q^2)$ and energies $(W)$ in terms of short and                               
long distance contributions.  They emphasize that even in the very low $Q^2$ region                              
the short distance contribution is not small, and also that at large $Q^2$ the long                              
distance effects still contribute.  Here we present a quantitative study which involves a                              
more precise approximation for the $q\bar{q} + p$ cross section and includes                               
consideration of the longitudinal structure function $F_L$.  Other differences are that                            
we compute the (small $k_T$) non-perturbative component using the VDM for small                            
$q\bar{q}$ masses $M < Q_0$ and the additive quark model for $M > Q_0$; we do                            
not need an artificial suppression\footnote{In ref.~\cite{GLM} an ad hoc suppression                            
factor of 0.6 is used.} of the VDM component.  Moreover we make a                               
detailed fit to the $F_2$ data in terms of an unintegrated gluon distribution which we                               
determine using an unified evolution equation which embodies both DGLAP and                               
BFKL evolution. \\                              
                              
\noindent{\large \bf 2.~The $\gamma^* p$ cross section}

The spectral function $\rho$ occurring in (\ref{eq:a1}) may be                                                
expressed in terms of  the $\gamma^* \rightarrow q \overline{q}$                                                
matrix element $\cal{M}$.  We have $\rho \: \propto | {\cal M}                                                
|^2$ with, for transversely polarised photons,                                                
\bea                                                
\label{eq:a9}                                                
{\cal M}_T & = & \frac{\sqrt{z (1 - z)}}{\overline{Q}^2 + k_T^2}                                                
\quad                                                
\overline{u}_\lambda (\gamma \cdot \varepsilon_\pm)                                                
u_{\lambda^\prime} \nonumber \\                                                 
\\                                                
& = & \frac{(\varepsilon_\pm \cdot k_T) [(1 - 2z) \lambda                                                
\mp 1] \: \delta_{\lambda, - \lambda^{\prime}} \; + \; \lambda                                                
m_q \; \delta_{\lambda \lambda^\prime}}{\overline{Q}^2 +                                                
k_T^2}. \nonumber \\ \nonumber                                                
\eea                                                
We use the notation of ref. \cite{LMRT}, which was based on the                                                
earlier work of ref. \cite{MBL}.  Namely the photon polarisation                                                
vectors are                                                
\be                                                
\label{eq:a10}                                                
\varepsilon_T \; = \; \varepsilon_\pm \; = \; \frac{1}{\sqrt{2}}                                                
\; (0, \; 0, \; 1, \; \pm i),                                                
\ee                                                
and $\lambda, \lambda^{\prime} = \pm \: 1$ corresponding to $q,                                                
\overline{q}$ helicities of $\pm \: \frac{1}{2}$.  Also we introduce                                               
\be                                                
\label{eq:a11}                                                
\overline{Q}^2 \; = \; z (1 - z) Q^2 + m_q^2.                                               
\ee                                                
Note that (\ref{eq:a9}) is written in terms of \lq\lq old-fashioned" time-ordered or                         
light cone perturbation theory where both the $q$ and $\bar{q}$ are on-mass-shell.                          
This form is appropriate when discussing the dispersion relation (\ref{eq:a1}) in the                         
$q\bar{q}$ invariant mass.  For high photon momentum $p_\gamma$ the two                         
time-ordered diagrams have a very different energy mismatch                        
\be                        
\label{eq:b11}                        
\left ( \Delta E \: \simeq \: \frac{Q^2 + M^2}{2p_\gamma} \right ) \; \ll \; \left ( \Delta                         
E^\prime \: \simeq \: p_\gamma \right ),                        
\ee                        
and so the contribution from the diagram $(\Delta E^\prime)$ with the \lq\lq wrong"                         
time-ordering may be neglected.  The remaining diagram, with energy denominator                         
$1/\Delta E$, leads to the behaviour $1/(\bar{Q}^2 + k_T^2)$ contained in                         
(\ref{eq:a9}), as can be seen on using (\ref{eq:a13}) below.                        
                        
In terms of the quark momentum variables $z, k_T^2$ of Fig.~1,                                                
equations (\ref{eq:a1}) and (\ref{eq:a2}) become                                                
\bea                                                
\label{eq:a12}                                                
\sigma_T & = & \sum_q \: \alpha \frac{e_q^2}{4 \pi^2} \; \sum_{\lambda \: =                                                
\: \pm \: 1} \: \int dz \: d^2 k_T \: ({\cal M}_T {\cal M}_T^*)                                                
N_c  \; \frac{1}{s} \; {\rm Im} \: A_{q \overline{q} + p}                                                
\nonumber \\                                                
\\                                                
& = & \sum_q \: \alpha \frac{e_q^2}{2 \pi} \; \int dz \: dk_T^2                                                
\frac{[z^2 + (1 - z)^2] k_T^2 + m_q^2}{(\overline{Q}^2 +                                                
k_T^2)^2} \; N_c \; \sigma_{q \overline{q} + p} (k_T^2) \nonumber                                                
\\ \nonumber                                                
\eea                                                
where the number of colours $N_c = 3$, and $e_q$ is the charge of                                                
the quark in units of $e$.  We shall give the corresponding cross section                                                
$\sigma_L$ for longitudinal polarised photons in the section~2.1.                                                
                                                
The dispersion relation (\ref{eq:a2}) in $M^2$ has become, in                                                
(\ref{eq:a12}), a two dimensional integral.  The relation between                                                
the variables is                                                 
\be                                                
\label{eq:a13}                                                
M^2 \; = \; \frac{k_T^2 + m_q^2}{z (1 - z)}                                                 
\ee                                                
where $m_q$ is the mass of the quark.  For massless quarks $z =                                                
\frac{1}{2} (1 + \cos \theta)$, where $\theta$ is the angle of the outgoing quark with                                                 
respect to the photon in the $q\bar{q}$ rest frame.  The $dz$ integration is implicit                                                
in (\ref{eq:a2}) as the integration over the quark angular                                                
distribution in the spectral function $\rho$.                                                
                                                
To determine $F_2 (x, Q^2)$ at low $Q^2$ we have to evaluate the contributions to                                                
$\sigma_T$ coming from the various kinematic domains.  First the contribution                                                
from the perturbative domain with $M^2 > Q_0^2$ and large $k_T^2$, and second                                                 
from the non-perturbative or long-distance domains.                                         
                                         
\medskip                                         
                                         
\noindent{\bf 2.1.~The $\gamma^* p$ cross section in the                                                
perturbative domain}                                                
                                                
We may begin with the two gluon exchange contribution to                                                 
quark--quark scattering                                                
\be                                                
\label{eq:a14}                                                
\sigma_{q + q} \; = \; \frac{2}{9} \; 4 \pi \: \int \alpha_S^2                                                
(l_T^2) \; \frac{dl_T^2}{l_T^4}                                                
\ee                                                
where $\pm \: \mbox{\boldmath$l$}_T$ are the transverse momenta of the gluons.                                                 
Thus                                                
for $q$-proton scattering we obtain                                               
  
\be                                                
\label{eq:a15}                                                
\sigma_{q + p} \; = \; \frac{2}{3} \; \pi^2 \int \alpha_S (l_T^2)                                                
\: f (x, l_T^2) \; \frac{d l_T^2}{l_T^4}                                                
\ee                                                
where                                                 
\be                                                
\label{eq:b15}                                                
f(x, l_T^2) = x \partial g (x, l_T^2) / \partial \ln                                                
l_T^2                                                
\ee                                                
is the unintegrated gluon density.  The process is shown in Fig.~2.                                                
Finally for $q \overline{q} +                                                
{\rm proton}$ scattering we have to include the graph for                                                
$\overline{q}                                                
+ p$ scattering. For both the $q$ and $\overline{q}$ interactions we                                                
have two diagrams of the type                                                
shown in Fig.~3 with ${\cal M}^* (\mbox{\boldmath$k$}_T +                                                
\mbox{\boldmath$l$}_T)$ and ${\cal M} (k_T)$.  We obtain                                                
\bea                                                
\label{eq:a16}                                                
\sigma_T & = & \sum_q \: \frac{\alpha e_q^2}{\pi} \;                                                
\int \: d^2 k_{1T} \: dz \: d^2 l_T \: \frac{f (x,                                                
l_T^2)}{l_T^4} \; \alpha_S (l_T^2) \; \times \nonumber \\                                                
\\                                                
&& \left \{ \left [ (1 - z)^2 + z^2 \right ] \: \left ( \frac{\mbox{\boldmath                            
$k$}_{1T}}{D_1} \: + \: \frac{\mbox{\boldmath $l$}_T - \mbox{\boldmath                            
$k$}_{1T}}{D_2} \right )^2 \: + \: m_q^2 \left ( \frac{1}{D_1} \: - \: \frac{1}{D_2}                            
\right )^2 \: \right \} \nonumber \\ \nonumber                                                
\eea                                                
where                                                
\be                                                
\label{eq:a17}                                                
x \; = \; (Q^2 + M^2) / s,                           
\ee                                                
\bea                           
D_1 & = & k_{1T}^2 \: + \: z (1 - z) Q^2 \: + \: m_q^2, \nonumber \\                           
& & \\                           
D_2 & = & (\mbox{\boldmath $l$}_T - \mbox{\boldmath $k$}_{1T})^2 \: + \: z (1 -                            
z) Q^2 \: + \: m_q^2. \nonumber                           
\label{eq:b17}                           
\eea                           
                        
Expression (\ref{eq:a16}) is written as the square of the amplitude for quark-antiquark                         
production, where we integrate over the quark momentum $k_{1T}$ in the inelastic                         
{\it intermediate} state, see Fig.~2.  The first term, proportional to $1/D_1$,                         
corresponds to the amplitude where the gluon couples to the antiquark $k_2$, while in                         
the second term, proportional to $1/D_2$, the gluon couples to the quark $k_1$.  Of                         
course form (\ref{eq:a16}) can also be used to calculate the cross section for high                         
$k_T$ dijet production $(\gamma^* p \rightarrow q\bar{q}p)$, where $k_{1T}$ and                         
$k_{2T}$ refer to the transverse momenta of the {\it outgoing} quark jets.                        
                        
To separate the perturbative and non-perturbative contributions to the cross section                         
(\ref{eq:a16}) for our inclusive process we have to introduce a cut on the quark                         
transverse momentum (as well as on the $q\bar{q}$ invariant mass $M$).  At first                         
sight it might appear that to obtain the perturbative contribution we simply require                         
$k_{1T} > k_0$.  However this implementation of the cut-off would not be correct.                          
For instance if, as in Fig.~2, the two exchanged gluons couple to the $k_1$ line, then                         
$\mbox{\boldmath $k$}_{2T} = \mbox{\boldmath $l$}_T - \mbox{\boldmath                         
$k$}_{1T}$ may be small and in the limit $m_q \rightarrow 0$ and small $Q^2$ we                         
would have an unphysical infrared singularity in the region of large $k_{1T}$ and                         
$l_T$, but small $k_{2T}$, coming from the $1/D_2$ term in (\ref{eq:a16}).  To see                         
better the origin of the infrared singularities we perform the square and write the                         
expression in curly brackets in (\ref{eq:a16}) in the form                        
\bea                   
\label{eq:c17}                        
\left \{ \frac{[(1 - z)^2 + z^2] k_{1T}^2 + m_q^2}{D_1^2} \right . & + & \frac{[(1 -             
z)^2 + z^2] (\mbox{\boldmath $l$}_T - \mbox{\boldmath $k$}_{1T})^2 + m_q^2}             
{D_2^2}  \nonumber \\                 
& & \\                
& + & 2 \left . \frac{[(1 - z)^2 + z^2] \mbox{\boldmath $k$}_{1T} \cdot                 
(\mbox{\boldmath $l$}_T -               
\mbox{\boldmath $k$}_{1T}) - m_q^2}{D_1 D_2} \right \}.  \nonumber                 
\eea                 
The danger comes from the second term, which corresponds to Fig.~2, whereas the                        
last term, which describes interference, is infrared stable, as we will show later.  Our                        
aim is to separate off all the infrared contributions into the non-perturbative part.                         
Therefore to evaluate the perturbative contribution coming from the second term we                        
have to use the cut-off $|\mbox{\boldmath $l$}_T - \mbox{\boldmath $k$}_{1T}| >                        
k_0$.  This is equivalent to changing the variable of integration for the second term                        
from $\mbox{\boldmath $k$}_{1T}$ to $\mbox{\boldmath $l$}_T -                        
\mbox{\boldmath $k$}_{1T}$, and so its contribution is exactly equal to that of the                        
first term.  An alternative way to introduce the same cut-off is to separate off the           
incoming $q\bar{q}$ configurations with $k_T < k_0$ so that (\ref{eq:a16}) becomes                       
\bea                       
\label{eq:d17}                       
\sigma_T & = & \sum_q \: \frac{2 \alpha e_q     ^2}{\pi} \: \int_{k_0^2} \: d^2 k_T dz                        
d^2 l_T \: \frac{f (x, l_T^2)}{l_T^4} \: \alpha_S (l_T^2) \nonumber \\                       
& & \\                       
& \times & \left \{ \frac{[(1 - z)^2 + z^2] k_T^2 + m_q^2}{(\bar{Q}^2 + k_T^2)^2}                        
\: - \: \frac{[(1 - z)^2 + z^2] \mbox{\boldmath $k$}_T \cdot (\mbox{\boldmath           
$k$}_T +                        
\mbox{\boldmath $l$}_T) + m_q^2}{(\bar{Q}^2 + k_T^2) (\bar{Q}^2 +                        
(\mbox{\boldmath $k$}_T + \mbox{\boldmath $l$}_T)^2)} \right \}. \nonumber                       
\eea                       
Note that the transverse momentum $\mbox{\boldmath $k$}_T$ of the incoming                        
quark is equal to $\mbox{\boldmath $k$}_{1T}$ when the gluon couples to the                        
antiquark (first term in (\ref{eq:c17})) and is equal to $\mbox{\boldmath $k$}_{1T} -                        
\mbox{\boldmath $l$}_{T}$ when the gluon couples to the quark (second term in                        
(\ref{eq:c17})).  Working in terms of the variable $\mbox{\boldmath $k$}_T$                        
corresponding to the dispersion cut shown in F 
ig.~1 has the advantage that it is then                        
easy to introduce cut-offs with respect to the invariant $q\bar{q}$ masses $M$ and                        
$M^\prime$, which we need to impose in order to separate off the non-perturbative           
VDM contribution\footnote{Of course the use of the Feynman rules would yield the           
same result, but the time-ordered or light cone approach with the incoming $q$ and                        
$\bar{q}$ on-shell is more convenient when we come to separate off the                        
non-perturbative component in terms of $k_T < k_0$ and $M, M^\prime < Q_0$.}.                       
                       
Another argument in the favour of the cut written in terms of initial quark momenta       
$k_T$ comes from the impact parameter representation.  Instead of $k_T$ we may       
use the transverse coordinate $b$ and write the cross section (\ref{eq:d17}) in the       
form      
\be      
\label{eq:c1}      
\sigma_T \; \propto \; \int dz d^2 b |\Psi_\gamma (b)|^2 f (x, b) \alpha_S (b)      
\ee      
where the gluon distribution       
\be      
\label{eq:c2}      
f (x, b) \; = \; \int \: \frac{d^2 l_T}{(2 \pi)^2}[1- e^{i \mbox{\boldmath $l$}_T \cdot  
\mbox{\boldmath $b$}}]  \frac{f (x, l_T^2)}{l_T^4}.      
\ee      
The photon \lq\lq wave function" is given by \cite{NZ2}      
\be      
\label{eq:c3}      
|\Psi_\gamma (b)|^2 \; = \; \sum_q \alpha e_q^2 [z^2 + (1 - z)^2] \bar{Q}^2 K_1^2       
(\bar{Q} b),     
\ee      
where for simplicity we have set $m_q = 0$.  The photon wave function is simply the       
Fourier transform of the matrix element ${\cal M}$ given by (\ref{eq:a9}).  It is most       
natural to take the infrared cut-off in coordinate space, say $b < b_0$.  The       
variable which is the Fourier conjugate of $b$ is the {\it incoming} quark       
momentum $k_T$ of Fig.~1 (rather than the {\it intermediate} transverse momentum      
$k_{1T}$ of Fig.~2).  This is further justification to impose the infrared cut in the      
form $k_T > k_0$.      
      
Now let us consider the interference contribution, that is the last term in           
(\ref{eq:d17}).  It is                        
infrared stable since in the limit $m_q^2 \rightarrow 0$ and $Q^2 \rightarrow 0$ it                        
takes the form                       
\be                       
\int \: \frac{d^2 k_T \mbox{\boldmath $k$}_T \cdot (\mbox{\boldmath $k$}_T +                        
\mbox{\boldmath $l$}_T)}{k_T^2 (\mbox{\boldmath $k$}_T + \mbox{\boldmath                        
$l$}_T)^2} \; \sim \; \int \: \frac{d (| \mbox{\boldmath $k$}_T + \mbox{\boldmath                        
$l$}_T|)}{k_T}                      
\label{eq:e17}                       
\ee                       
when $|\mbox{\boldmath $k$}_T + \mbox{\boldmath $l$}_T|$ is small. We have                        
used boundaries $k_T^2 = k_0^2$ and $M^2 = Q_0^2$                                                 
to separate the perturbative QCD (pQCD), additive quark model                                                
(AQM) and vector meson dominance (VDM) contributions.  As a result the                                                 
$\gamma^* p$ cross section formulae, (\ref{eq:d17}), is                                                 
asymmetric between the ingoing and outgoing quarks.                                                
The origin of the asymmetry is the difference of the transverse                                                
momentum of the outgoing quark $(\mbox{\boldmath$k$}_T +                                                 
\mbox{\boldmath$l$}_T)$                                                
and the incoming quark $(\mbox{\boldmath$k$}_T)$ in Fig.~3.  Such a graph                                                
therefore represents the interference between $M$ and $M^\prime \neq M$ states.                                                
To obtain the pure pQCD contribution we require the incoming $q \overline{q}$                        
system to satisfy $M^2 > Q_0^2$ and $k_T > k_0$.  Ideally we would like to impose                                                
the same cuts on the outgoing $q \overline{q}$ system, namely                                                
\be                                                
\label{eq:z2}                                                
M^{\prime 2} \; = \; \frac{(\mbox{\boldmath$k$}_T + \mbox{\boldmath$l$}_T)^2 \:                                                
+ \: m_q^2}{z (1 - z)} \; > \; Q_0^2                                                
\ee                                                
and $k_T^\prime = | \mbox{\boldmath$k$}_T + \mbox{\boldmath$l$}_T | > k_0$.                                                
However in a small region of phase space, where $\mbox{\boldmath$l$}_T$ lies                                                 
close                                                
to $- \mbox{\boldmath$k$}_T$, we may have $M^\prime < Q_0$ and/or                                                 
$k_T^\prime < k_0$.  For this region we therefore have interference between the                                                
pQCD and VDM (or AQM) contributions.  There is no double counting since neither                                                
our VDM or AQM\footnote{For the AQM contribution the interaction with the target                                                 
proton is described                                                
by the forward elastic quark scattering amplitude and hence we have $z^\prime = z, \;                                                
k_T^\prime = k_T$ and $M = M^\prime$.} components contain interference terms.                                                
This is fortunate because we cannot neglect the contribution from this small                                                
part of phase space of Fig.~3 without destroying gauge invariance, which is                                                
provided by the sum of the graphs in Figs.~2 and 3.  We stress that the                                                
contribution coming from this limited region $\mbox{\boldmath$l$}_T$ close to                                                
$-\mbox{\boldmath$k$}_T$ is infrared stable and hence it is small and has little                        
impact on the overall fit to the data.                                                
                       
So far we have only calculated $\sigma_T$.  In the same way we may calculate the                        
cross section for                                                
longitudinally polarised incident photons.  In this case the                                                
relation analogous to (\ref{eq:a12}) reads                                                
\be                                                
\label{eq:a18}                                                
\sigma_L \; = \; \sum_q \: \frac{\alpha e_q^2}{2 \pi} \; \int dz dk_T^2 \;                                                
\frac{4 Q^2 \: z^2 (1 - z)^2 N_c}{(\overline{Q}^2 + k_T^2)^2} \;                                                
\sigma_{q \overline{q} + p} (k_T^2),                                                
\ee                                                
which on evaluating $\sigma_{q \overline{q} + p}$ gives                                                
\bea                                                
\label{eq:a19}                                                
\sigma_L & = & \sum_q \: \frac{2 \alpha e_q^2}{\pi} \; Q^2                                                
\int_{k_0^2} \: d^2 k_T dz \: d^2 l_T \; \frac{f                                                
(x, l_T^2)}{l_T^4} \: \alpha_S (l_T^2) 4z^2 (1 - z)^2 \; \times \nonumber \\                        
& & \\                        
& & \left \{ \frac{1}{(\bar{Q}^2 + k_T^2)^2} \; - \; \frac{1}{(\bar{Q}^2 + k_T^2)                        
(\bar{Q}^2 + (\mbox{\boldmath $k$}_T + \mbox{\boldmath $l$}_T)^2 )} \right \}.                        
\nonumber                        
\eea                        
                             
>From the formal point of view the integrals over $l_T^2$ and                                                
$k_T^2$ cover the interval 0 to $\infty$.  For the $l_T^2$                                                
integration in the domain $l_T^2 < l_0^2 \sim 1 {\rm GeV}^2$ we                                                
may use the approximation                                                
\be                                                
\label{eq:a20}                                                
\alpha_S (l_T^2) \: f (x, l_T^2) \; = \; \frac{l_T^2}{l_0^2} \: \alpha_S (l_0^2)                                                
\; f (x, l_0^2).                                                
\ee                                                
For $k_T^2 < k_0^2$ we enter the long distance domain which we                                                
discuss next.  To be precise we use the formula (\ref{eq:d17})                                                
and (\ref{eq:a19}) to evaluate the cross sections only in the                                                
perturbative domain $M^2 > Q_0^2$ and $k_T^2 > k_0^2$.  We                                                
exclude the region $M^2 < Q_0^2$ and $k_T^2 > k_0^2$ from the                                                
perturbative domain as the point-like (short-distance) component                                                
of the vector meson wave function will be included in the VDM                                                
term.                                         
                                         
\medskip                                         
                                                
\noindent{\bf 2.2.~The $\gamma^* p$ cross section in the                                                
non-perturbative domain}                                                
                                                
There are two different non-perturbative contributions.  First                                                
for $M^2 < Q_0^2$ we use the conventional vector meson dominance                                                
formula (\ref{eq:a7}) for $F_T (x, Q^2)$.  We also should include                                                
the longitudinal structure function $F_L (x, Q^2)$.  $F_L$ is                                                
given by a formula just like (\ref{eq:a7}) but with the                                                
introduction of an extra                                                
factor $\xi Q^2/M_V^2$ on the right-hand side.  $\xi (Q^2)$ is a                                                
phenomenological function which should decrease with increasing                                                
$Q^2$.  The data for $\rho$ production indicate that $\xi                                                
(m_\rho^2) \lapproxeq 0.7$ \cite{EE}, whereas at large $Q^2$ the                                                
usual properties of deep inelastic scattering predict that                                                 
\be                                                
\label{eq:a21}                                                
\frac{F_L}{F_T} \; \sim \; \frac{4 k_T^2}{Q^2} \; \lapproxeq \;                                                
\frac{M_V^2}{Q^2}.                                                
\ee                                                
So throughout the whole $Q^2$ region the contribution of $F_L$ is                                                
less than that of $F_T$.  In order to calculate $F_L$ (VDM) we                                                
insert the factor $\xi Q^2 / M_V^2$ in (\ref{eq:a7}) and use an                                                
interpolating formula for $\xi$                                                
\be                                                
\label{eq:a22}                                                
\xi \; = \; \xi_0 \left(\frac{M_V^2}{M_V^2 + Q^2} \right)^2                                                
\ee                                                
with $\xi_0 = 0.7$, which accommodates both the $\rho$ meson                                                
results and the deep inelastic expectations of (\ref{eq:a21}).  However the recent                                     
$\rho$ electroproduction, $\gamma^* p \rightarrow \rho p$, measurements                                     
\cite{RHO} indicate that $\sigma_L (\rho)/\sigma_T (\rho)$ may tend to a constant                                     
value for large $Q^2$.  We therefore also show the effect of calculating $F_L$                                     
(VDM) from (\ref{eq:a7}) using                                    
\be                                    
\label{eq:b22}                                    
\xi \; = \; \xi_0 \: \left ( \frac{M_V^2}{M_V^2 + Q^2} \right ),                                    
\ee                                    
see Fig.~9 below.                                    
                                                
The second non-perturbative contribution covers the low $k_T$ part of                                                
the $M^2 > Q_0^2$ domain, that is the region with $k_T^2 < k_0^2$.  Here we                                                 
use the additive quark model                                                
and the impulse approximation to evaluate the $\sigma_{q                                                
\overline{q} + p}$ cross sections in formulas (\ref{eq:a12}) and                                                
(\ref{eq:a18}).                                         
                                         
\medskip                                         
                                                
\noindent{\bf 2.3.~Final formulae}                                                
                                                
For completeness we list below the formulae that we use for the non-pQCD                                                 
contributions coming from the $k_T < k_0$ domain. When $M < Q_0$, with $Q_0^2                                                
\simeq 1 - 1.5 {\rm GeV}^2$, we use the vector meson dominance model.  We have                                                
\be                                                
\label{eq:z5}                                                
\sigma_T ({\rm VDM}) \; =\; \pi \alpha \sum_{V = \rho, \omega, \phi} \;                                                
\frac{M_V^4 \; \sigma_V (W^2)}{\gamma_V^2 (Q^2 + M_V^2)^2}                                                
\ee                                                
\be                                                
\label{eq:z6}                                                
\sigma_L ({\rm VDM}) \; = \; \pi \alpha \sum_{V = \rho, \omega, \phi} \;                                                
\frac{Q^2 M_V^2 \; \sigma_V (W^2)}{\gamma_V^2 (Q^2 + M_V^2)^2} \; \xi_0 \;                                                
\left( \frac{M_V^2}{Q^2 + M_V^2} \right)^2                                                
\ee                                                
with $\xi_0 = 0.7$, see (\ref{eq:a22}).  For the vector meson-proton cross                                                 
sections, we take                                                
\bea                                                
\label{eq:z9}                                                
\sigma_\rho & = & \sigma_\omega \; = \; \textstyle{\frac{1}{2}}                                                 
\left[ \sigma (\pi^+ p) + \sigma (\pi^- p) \right] \nonumber \\                                                
\sigma_\phi & = & \sigma (K^+ p) + \sigma (K^- p) - \textstyle{\frac{1}{2}}                                                 
\left[\sigma (\pi^+ p) + \sigma (\pi^- p) \right].                                               
\eea                                                
Finally for $M > Q_0$ (and $k_T < k_0$)                                                
we use the additive qua 
rk model and impulse approximation                                                
\be                                                
\label{eq:z7}                                                
\sigma_T ({\rm AQM}) \; = \; \alpha \sum_q \; \frac{e_q^2}{2 \pi} \; \int \:                                                
d z dk_T^2 \; \frac{[z^2 + (1 - z)^2] k_T^2 + m_q^2}{(\tilde{Q}^2 + k_T^2)^2} \;                                                
N_c \: \sigma_{q \overline{q} + p} \; (W^2)                                                
\ee                                                
\be                                                
\label{eq:z8}                                                
\sigma_L ({\rm AQM}) \; = \; \alpha \sum_q \; \frac{e_q^2}{2 \pi} \; \int \:                                                
d z d k_T^2 \; \frac{4 Q^2 \; z^2 (1 - z)^2}{(\tilde{Q}^2 + k_T^2)^2} \; N_c                                                
\: \sigma_{q \overline{q} + p} \; (W^2)                                                
\ee                                                
where for $\sigma_{q \overline{q} + p}$ we take, for the light quarks,                                                
\be                                                
\label{eq:z10}                                                
\sigma_{q \overline{q} + p} \: (W^2) \; = \; \frac{2}{3} \; \sigma_{p                                                 
\overline{p}} \: ( s = \textstyle{\frac{3}{2}} W^2).                                                
\ee                                                
The \lq \lq photon" wave function contains propagators like $1/(\overline{Q}^2                                                
+ k_T^2)$ and in impact parameter $b_T$ space it receives contributions from                                                 
the whole of the $b_T$ plane extending out to infinity.  On the other hand                                                 
confinement restricts the quarks to have limited separation, say                                                
$b_T = | \mbox{\boldmath$b$}_{1T} - \mbox{\boldmath$b$}_{2T}| \lapproxeq                                                 
1$~fm.  To allow for this effect we have replaced $\overline{Q}^2$ by $\tilde{Q}^2                                                 
= \overline{Q}^2 + \mu^2$ in (\ref{eq:z7}) and (\ref{eq:z8}), where $\mu$ is                                                 
typically the inverse pion radius.  We therefore take $\mu^2 = 0.1                                                 
{\rm GeV}^2$.  This change has no effect for $Q^2 \gg \mu^2$ but for $Q^2                                                 
\lapproxeq \mu^2$ it gives some suppression of the AQM contribution.                                          
                                         
\medskip                                         
                                                
\noindent{\bf 2.4.~The quark mass}                                                
                                                
In the perturbative QCD domain we use the (small) current quark                                                
mass $m_{{\rm curr}}$, while for the long distance contributions                                                
it is more natural to use the constituent quark mass $M_0$.  To                                                
provide a smooth transition between these values (in both the AQM and perturbative                            
QCD domains) we take the running mass obtained                                                 
from a QCD-motivated model of the spontaneous chiral symmetry breaking in the                                                 
instanton vacuum \cite{INST}                                                
\be                                                
\label{eq:a23}                                                
m_q^2 \; = \; M_0^2 \: \left ( \frac{\Lambda^2}{\Lambda^2 + 2 \mu^2} \right )^6 \: +                                                 
\: m_{\rm curr.}^2.                                                
\ee                                                
The parameter $\Lambda = 6^{1/3}/\rho = 1.09$~GeV, where $\rho = 1/(0.6~{\rm                                                 
GeV})$ is the typical size of the instanton.  $\mu$ is the natural scale of the problem,                                                 
that is $\mu^2 = z (1-z) Q^2 + k_T^2$ or $\mu^2 = z (1-z) Q^2 + (\mbox{\boldmath                                                 
$l$}_T + \mbox{\boldmath $k$}_T)^2$ as appropriate.  For constituent and current                            
quark masses we take $M_0 = 0.35$~GeV and $m_{\rm curr} = 0$ for the $u$ and                            
$d$ quarks, and $M_0 = 0.5$~GeV and $m_{\rm curr} = 0.15$~GeV for the $s$                            
quarks.  \\                                                
                                             
\noindent{\large \bf 3.~The description of the data for $F_2$}                                                
                                               
Though in principle it would appear that we have a parameter-free\footnote{Apart of                                             
course from the form of the input gluon distribution, $g (x, l_0^2$).} prediction of                                             
$F_2                                                
(x, Q^2)$ at low $Q^2$, in practise we have to fix the values of the parameters                                                
$k_0^2$ and $Q_0^2$.  Recall that $k_T^2 = k_0^2$ specifies the boundary between                                                
the non-perturbative and perturbative QCD components, and that $M^2 = Q_0^2$                                                
specifies the boundary between the VDM and AQM contributions to the non-                                               
perturbative component.  The results that we present correspond to the choice $Q_0^2                                                
= 1.5$~GeV$^2$, for which the VDM contribution is computed from the $\rho,                                                
\omega$ and $\phi$ meson contributions (with mass $M_V < Q_0$).  The more                                                
sensitive parameter is $k_0^2$.  We therefore present results for two choices, namely                                                
$k_0^2 = 0.2$ and 0.5~GeV$^2$, which show some interesting and observable                                                
differences.  The results are much more stable to the increase of $k_0^2$ from 0.5 to                                                
1~GeV$^2$.                                               
                                               
To calculate the perturbative contributions we need to know the unintegrated gluon                                                
distribution $f (x, l_T^2)$, see (\ref{eq:d17}) and (\ref{eq:a19}).  To determine $f                                                
(x, l_T^2)$ we carry out the full programme described in detail in                                                
Ref.~\cite{KMS}.  We solve a \lq\lq unified\rq\rq~equation for $f (x, l_T^2)$                                                
which incorporates\footnote{Following Ref.~\cite{KMS} we appropriately constrain                                                
the transverse momenta of the emitted gluons along the BFKL ladder.  There is                                                
an indication, from comparing the size of the next-to-leading $\ln (1/x)$ contribution                                               
\cite{CCFL} to the BFKL intercept with the effect due to the kinematic constraint                                               
\cite{KMS1}, that the incorporation of the constraint into the evolution analysis gives                                               
a major part of the subleading $\ln (1/x)$ corrections.} BFKL and DGLAP evolution                                               
on an equal footing, and al 
lows the description of both small and large $x$ data.  To                                               
be precise we solve a coupled pair of integral equations for the gluon and sea quark                                                
distributions, as well as allowing for the effects of valence quarks.  As in                            
Ref.~\cite{KMS} we take $l_0^2 = 1$~GeV$^2$, but due to the large anomalous                            
dimension of the gluon the results are quite insensitive to the choice of $l_0$ in the                            
interval 0.8--1.5~GeV.                           
                                               
The starting distributions for the evolution are specified in terms of three parameters                                                
$N, \lambda$ and $\beta$ of the gluon                                               
\be                                               
\label{eq:a25}                                               
xg (x, l_0^2) \; = \; N x^{- \lambda} (1 - x)^\beta                                               
\ee                                               
where $l_0^2 = 1$~GeV$^2$.  At small $x$ the gluon drives the sea quark                                               
distribution.  The $k_T$ factorization                                                
theorem gives                                               
\be                                               
\label{eq:a26}                                               
S_q (x, Q^2) \; = \; \int_x^1 \: \frac{dz}{z} \: \int \: \frac{dk^2}{k^2} \: S_{\rm                                                
box}^q  (z, k^2, Q^2) \: f \left ( \frac{x}{z}, Q^2 \right )                                               
\ee                                               
where $S_{\rm box}$ describes the quark box (and crossed box) contribution.  The                                                
full expression for $S_{\rm box}$ is given in Ref.~\cite{KMS}.  Thus the sea $S_q$                                                
is given in terms of the gluon $f$ except for the contribution from the                                                
non-perturbative region $k^2 < k_0^2$, where we take                                               
\be                                               
\label{eq:a27}                                               
S_u^{\rm np} \; = \; S_d^{\rm np} \; = \; 2S_s^{\rm np} \; = \; C \: x^{-0.08} (1 -                                                
x)^8.                                               
\ee                                               
The parameter $C$ is fixed by the momentum sum rule in terms of the parameters $N,                                                
\lambda$ and $\beta$ specifying the gluon.  The charm component of the sea is                                                
obtained entirely from perturbative QCD (see \cite{KMS}) with the charm mass $m_c                                  
= 1.4$~GeV.  The valence quark contribution plays a very minor role in our analysis                                  
and so we take it from the GRV set\footnote{The GRV valence distributions were                                  
fitted to the MRS(A) distributions \cite{MRSA} at $Q^2 = 4$~GeV$^2$.} of partons                                  
\cite{GRV}.  Of course the sea quark distributions $S_q (x, Q^2)$ of                                               
(\ref{eq:a26}) (and (\ref{eq:a27})) are used only to get a more precise determination                                               
of $f (x, l_T^2)$ through the coupled evolution equations.  These forms for $S_q$                                             
are not used in our fit to the $F_2$ data since the sea contribution is already                                             
embedded in (\ref{eq:d17}) and (\ref{eq:a19}).                                              
                                               
\begin{table}[t]                                              
\caption{The values of the gluon parameters of eq.~(\protect \ref{eq:a25}).}                                            
\begin{center}                                               
\begin{tabular}{|cc|cccc|} \hline                                               
& $k_0^2$ & $N$ & $\lambda$ & $\beta$ & $\chi^2$/datapoint \\                                               
& (GeV$^2$) & & & & [423 points] \\ \hline                                               
Fit A & 0.2 & 0.97 & 0.16 & 3.6 & 1.09 \\                                               
Fit B & 0.5 & 0.42 & 0.32 & 3.7 & 1.70 \\ \hline                                               
\end{tabular}                                               
\end{center}                                              
\end{table}                                              
                                               
We determine the parameters $N, \lambda$ and $\beta$ by fitting to the available data                                                
for $F_2$ with $x < 0.05$.  We present two fits corresponding to a larger perturbative                                                
QCD contribution (Fit A with $k_0^2 = 0.2$~GeV$^2$) and a smaller pQCD                                                
component (Fit B with $k_0^2 = 0.5$~GeV$^2$).  The values of the gluon                                                
parameters are given in Table I and the quality of the description of the $F_2$ data is                                                
shown in Fig.~4.  Only a selection of the data fitted are shown in Fig.~4.  Both                                                
descriptions are in general satisfactory, but Fit A is superior mainly due to Fit B                                                
lying below the data for $Q^2 \simeq 1$~GeV$^2$.  This difference is better seen in                                                
Fig.~5 which shows the fit as a function of $Q^2$ for various fixed values of $x$.                                                
We see that Fit A, with the larger perturbative component is more able to                                               
accommodate the charge in slope going from high to low $Q^2$.  It is informative to                                               
show the components of the cross section.  The breakdown is shown in Figs.~6 and 7                                               
for Fits A and B respectively for the maximum energy $W = 245$~GeV for which                                               
data are available.  It appears that the low $Q^2$ behaviour of the pQCD component                                               
with low $l_T$ plays the vital role.                                            
                                            
The description of the $F_2$ data by Fit A is better than that obtained by Badelek and                                             
Kwiecinski \cite{BK}, which is to be expected since we perform a fit to the data,                                             
albeit with a very economical parametrization. Fig.~5 also shows the HERA                                  
photoproduction measurements                                     
at $W = 170$ and 210~GeV.  These data are not included in the fit.  We see that our                                     
description overshoots the published H1 \cite{H1P} and ZEUS \cite{ZP}                                    
measurements, although by a smaller margin than that of ref.~\cite{BK}.  On the other                                  
hand our extrapolation is in excellent agreement with a subsequent analysis of ZEUS  
data performed in ref.~\cite{SU}.  We will return to the comparison with                                    
photoproduction data when we study the effects of a different choice of the quark                                    
mass. \\                                    
                                            
\noindent {\large \bf 4.  Discussion}

We have made what appears to be in principle a prediction of $F_2$, or rather of                                             
$\sigma_{\gamma^* p}$, over the entire $Q^2$ range which relies only on the form                                             
of the initial gluon distribution, see (\ref{eq:a25}) and the parameter values of                                             
Table~1.  However a comparison of the results of Fits A and B show that in practice                                             
the results are dependent on the choice of the boundary $k_T^2 = k_0^2$ between the                                             
perturbative and non-perturbative contributions, where $\pm \mbox{\boldmath                                             
$k$}_T$ are the transverse momenta of the incoming $q$ and $\bar{q}$ which result                                             
from the $\gamma^* \rightarrow q\bar{q}$ transition.                                            
                                            
There are compelling reasons to select Fit A with $k_0^2 = 0.2$~GeV$^2$, which has                                             
the larger perturbative QCD contribution.  Fit A is not                                             
only preferred by the data, but it also yields an input gluon with a more reasonable                                             
small $x$ behaviour.  In fact for Fit A $(k_0^2 = 0.2$~GeV$^2$) the AQM                                             
contribution is almost negligible and the fit produces a reasonable $\lambda$, namely                                             
$\lambda = 0.16$.  On the other hand Fit B (with $k_0^2 = 0.5$~GeV$^2$) requires a                                             
larger $\lambda$, $\lambda = 0.32$, in order to compensate for the much more flat                            
$x^{-0.08}$ behaviour of the rather large AQM component.  Further support for Fit A                            
comes from the predictions for the longitudinal structure function, $F_L$.  Fig.~8                            
shows that the prediction from Fit B is much larger than that of Fit A due mainly to                            
the large AQM contribution.  Fig.~8 also shows the expectations for $F_L$ from the                            
analysis of ref.~\cite{BKS} and from the MRST partons \cite{MRST} of the most                            
recent global parton analysis.  We see these independent determinations of $F_L$                            
favour the prediction of Fit A.                                   
                                  
For completeness we show by the dashed curve in Fig.~9 the predictions of                                   
$\sigma_L/\sigma_T$ versus $Q^2$ obtained from Fit A.  This figure also shows the                                   
effect of replacing (\ref{eq:a22}) by (\ref{eq:b22}) in the formula for the VDM                                   
contribution to $F_L$.  Recall that (\ref{eq:b22}) was motivated by the possibility                                  
that the ratio $\sigma_L (\rho)/\sigma_L (\rho)$ for $\rho$ meson electroproduction                                  
tends to a constant value $A$ as $Q^2 \rightarrow \infty$.  We see from Fig.~9 that                                  
this change to the VDM contribution affects $F_L$, and hence $\sigma_L/\sigma_T$,                                   
mainly in the interval $0.2 < Q^2 < 10$~GeV$^2$.  It is straightforward to deduce                                   
from Fig.~9 the effect of changing the value of the parameter $\xi_0$ of                                  
(\ref{eq:b22}) to match the constant limit $A$ observed for the $\rho$ ratio.                                   
                                               
A remarkable feature of the recent measurements \cite{H1,ZEUS} of                                                
$\sigma (\gamma^* p) \: = \: (4 \pi^2 \alpha / Q^2) \: F_2 (x, Q^2)$ at fixed $W$, is                                  
the transition from a flat behaviour in the low $Q^2$ domain to the steep $\sim \:                                  
Q^{-2}$ fall off characteristic of perturbative QCD, see Fig.~5.  The                                                
transition appears to occur at $Q^2 \sim 0.2 {\rm GeV}^2$.  Such a break with                                  
decreasing $Q^2$ may reflect either the saturation due to the onset of absorption                                  
corrections or the fact that we are entering the confinement domain.  The observed                                                
features of the data favour the last possibility.  First there is no similar break in the                                  
behaviour of $F_2$ as a function of $x$ at low $x$ which would be expected if                                  
absorptive corrections were important.  A related observation is that the break, as a                                                
function of $Q^2$, appears to occur at the same value $Q^2 \sim 0.2 {\rm GeV}^2$                                  
for those $W$ values for which data are available.  Moreover we directly estimated                                  
the effect of the absorptive corrections using the eikonal rescattering model and found                                  
that they give a negligibly small effect on the $Q^2$ behaviour of the cross section                                  
and of $F_2$.  On the other hand, if the break is due to confinement then it is                                  
expected to occur at a value of $\overline{Q}^2$ which corresponds to the distances                                  
of the order of 1 fm, that is                                                 
\be                                                
\label{eq:b1}                                                
z (1 - z) \: Q^2 \; \sim \; Q^2/5 \; \sim \; (0.2 {\rm GeV})^2                                                
\ee                                                
which gives $Q^2 \sim 0.2 {\rm GeV}^2$ where the break is                                                
observed.                                                
                                                
In our calculations we have used a running quark mass which links                                             
the current $(m_{{\rm curr}})$ to the constituent $(M_0)$ mass.  The growth of                                             
$m_q$ in the transition region from perturbative QCD to the large distance domain is                                             
an important non-perturbative effect, which we find is required  by the $F_2$ data.                                             
>From the theoretical point of view such a behaviour of $m_q$ may be generated by                                             
the spontaneous breakdown of chiral symmetry in the instanton QCD vacuum                                                
\cite{INST}.  The qualitative features are that $m_q \sim M_0$ if the virtuality $q^2$                                             
of the quark is less or of the order of the square of the inverse of the instanton size, but                                             
that $m_q$ decreases quickly as $q^2$ increases.  In our analysis we have used a                                                
simplified power approximation for $m_q$, see (\ref{eq:a23}).                                               
                              
It is interesting to explore the effect of a different choice of quark mass.  The dashed                               
curves in Fig.~10 show the effect of using the constituent (fixed) mass $M_0$ of the                               
quarks in all the contributions to $F_2$ or $\sigma (\gamma^* p)$.  As expected in                              
the large $Q^2 \gg M_0^2$ perturbative domain the change has little effect.  For small                               
$Q^2$ it reduces the predictions.  For example, the photop 
roduction estimates for $W                               
\sim 200$~GeV are reduced by more than 10\% and would bring our analysis more                               
into line with the published H1 and ZEUS photoproduction measurements.  However                              
our running                               
quark mass predictions (continuous curves) are more physically motivated and should                               
be more reliable.  It will be interesting to see if their agreement with the experimental  
values extracted in ref.~\cite{SU} is maintained when the new photoproduction  
measurements are available from the HERA experiments.                               
                                                
A noteworthy point of our description of the $F_2$ data is the                                                
importance of the non-diagonal $(M \neq M^{\prime})$ perturbative                                                
QCD contribution to the double dispersion relation (\ref{eq:a1}).                                                
The contribution, which comes from the interference terms in (\ref{eq:d17}) (and        
(\ref{eq:a19})), corresponds to the diagram shown in Fig.~3.  It clearly has a negative      
sign, and moreover                                                
\be                                                
\label{eq:a24}                                                
\left\{M^2 \; = \; \frac{k_T^2 + m_q^2}{z (1 - z)} \right\} \;                                                
\neq \; \left\{M^{\prime 2} \; = \; \frac{(\mbox{\boldmath$k$}_T                                                
+ \mbox{\boldmath$l$}_T)^2 + m_q^2}{z (1 - z)} \right\}.                                                
\ee                                                
After the integration over the azimuthal angle in (\ref{eq:d17}), the interference term       
exactly cancels the diagonal first term for any $l_T < k_T$ in the limit of $Q^2       
\rightarrow 0$ and $m_q = 0$.  As a result the perturbative component of the cross       
section coming from the region of small $l_T$ essentially vanishes\footnote{Of       
course there is also a non-negligible contribution coming from the domain $l_T >       
k_T$ which does not vanish as $Q^2 \rightarrow 0$.} as $Q^2 \rightarrow 0$.  This                                                
property, seen in the $l_T < l_0$ components shown in Figs. 6 and 7, helps to                                                
reproduce the very flat $Q^2$ behaviour of $\sigma (\gamma^* p)$                                                
observed at low $Q^2, \; Q^2 \lapproxeq 0.2 {\rm GeV}^2$.  The fact that this low                               
$l_T$ gluon contribution becomes very small as $Q^2$ decreases (and in fact       
vanishes for $l_T < k_T$ in the $Q^2 \rightarrow 0$ limit) may be                               
considered as a justification of the perturbative QCD contribution to $F_2$ for low                               
$Q^2$.  The VDM cross section (and other diagonal contributions as well)                                                
decrease as $1 / (M_V^2 + Q^2)^2$ so we require just such a                                                
component which increases with $Q^2$ in order to compensate the                                                
decrease of the diagonal terms.  The compensation is well                                                
illustrated by Figs.~6 and 7 which show the behaviour of the various                                                
components as a function of $Q^2$.  Of course the compensation (that is the effect of       
the vanishing of the low $l_T^2$ contribution as $Q^2 \rightarrow 0$) is more       
manifest in the Fit A where a larger part of the phase space is described in terms of       
perturbative QCD.        
      
It is interesting to note that in this paper we have included two different types of       
interference effect.  First we have the dominant interference between the large $M$       
and $M^\prime$ states which gives rise to the decrease of the pure perturbative small      
$l_T$       
component of the cross section as $Q^2 \rightarrow 0$, and which is responsible for       
the good description of the low $Q^2$ data.  Then there is the interference between       
the perturbative and non-perturbative amplitudes which we have modelled using the       
perturbative formula in the region of small $M^\prime$ and/or small       
$|\mbox{\boldmath $k$}_T + \mbox{\boldmath $l$}_T|$.  We have noted that this       
contribution is small due to the infrared stability of the integral, as was shown in      
(\ref{eq:e17}).     
     
In summary we obtain an excellent description of $F_2$, or rather of      
$\sigma_{\gamma^* p}$, over the entire $Q^2$ range (from very low to high values                                           
of $Q^2$) in terms of physically motivated perturbative and non-perturbative                                             
contributions.  The choice of the boundary between the perturbative and                                             
non-perturbative domains which gives an excellent fit to the data, is also found to                                             
yield a sensible gluon distribution and reasonable predictions for $F_L$. \\                                            

\newpage        
\noindent{\large \bf Acknowledgements}                                                
                                                
We thank Krzysztof Golec-Biernat, Jan Kwiecinski and Vladimir Shekelyan                              
for valuable discussions and their interest in this work.                              
 MGR thanks the Royal Society, INTAS (95-311) and the Russian Fund of                             
Fundamental                                                 
Research (98~02~17629), for support.  AMS thanks the Polish State Committee for                                          
Scientific Research (KBN) grants No.~2 P03B~089~13 and 2~P03B~137~14 for                                          
support.  Also this work was supported in part by the EU Fourth Framework                                          
Programme \lq Training and Mobility of Researchers', Network \lq Quantum                                          
Chromodynamics and the Deep Structure of Elementary Particles', contract                                          
FMRX-CT98-0194 (DG~12 - MIHT).

\newpage                                               
                                                
\noindent{\large \bf Figure Captions} \\                                                
\begin{itemize}                                                
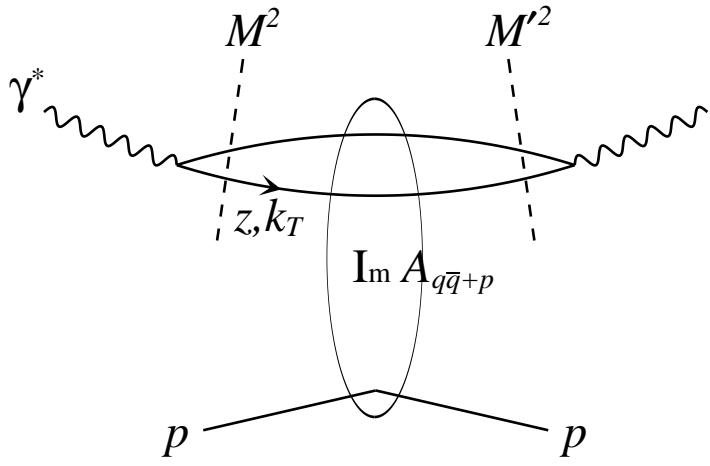
\item[Fig.~1]  The schematic  representation of the double                                                
dispersion (\ref{eq:a1}) for the $\gamma^* p$ total cross section                                                
$\sigma (s, Q^2)$ at fixed c.m. energy $\sqrt{s}$.  The cut                                                
variables, $M$ and $M^{\prime}$, are the invariant masses of the                                                
incoming and outgoing $q \overline{q}$ states in the                                                
quasi-elastic forward amplitude, $A_{q \overline{q} + p}$.                                                
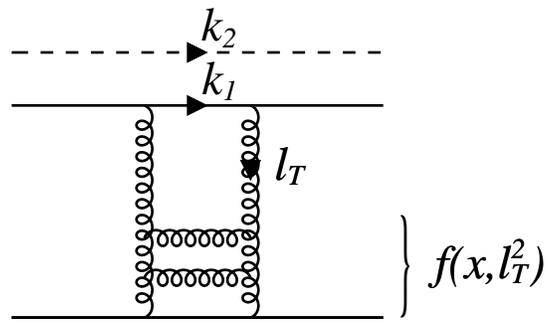
\item[Fig.~2]  The quark-proton interaction via two gluon                                                
exchange.  The spectator (anti)quark is shown by the dashed                                                
line.  $f (x, l_T^2)$ is the unintegrated gluon distribution of the proton.                                                
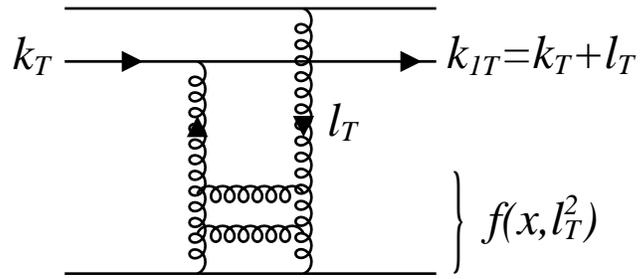
\item[Fig.~3]  A \lq \lq non-diagonal" $q \overline{q} - {\rm                                                
proton}$ interaction.                                                
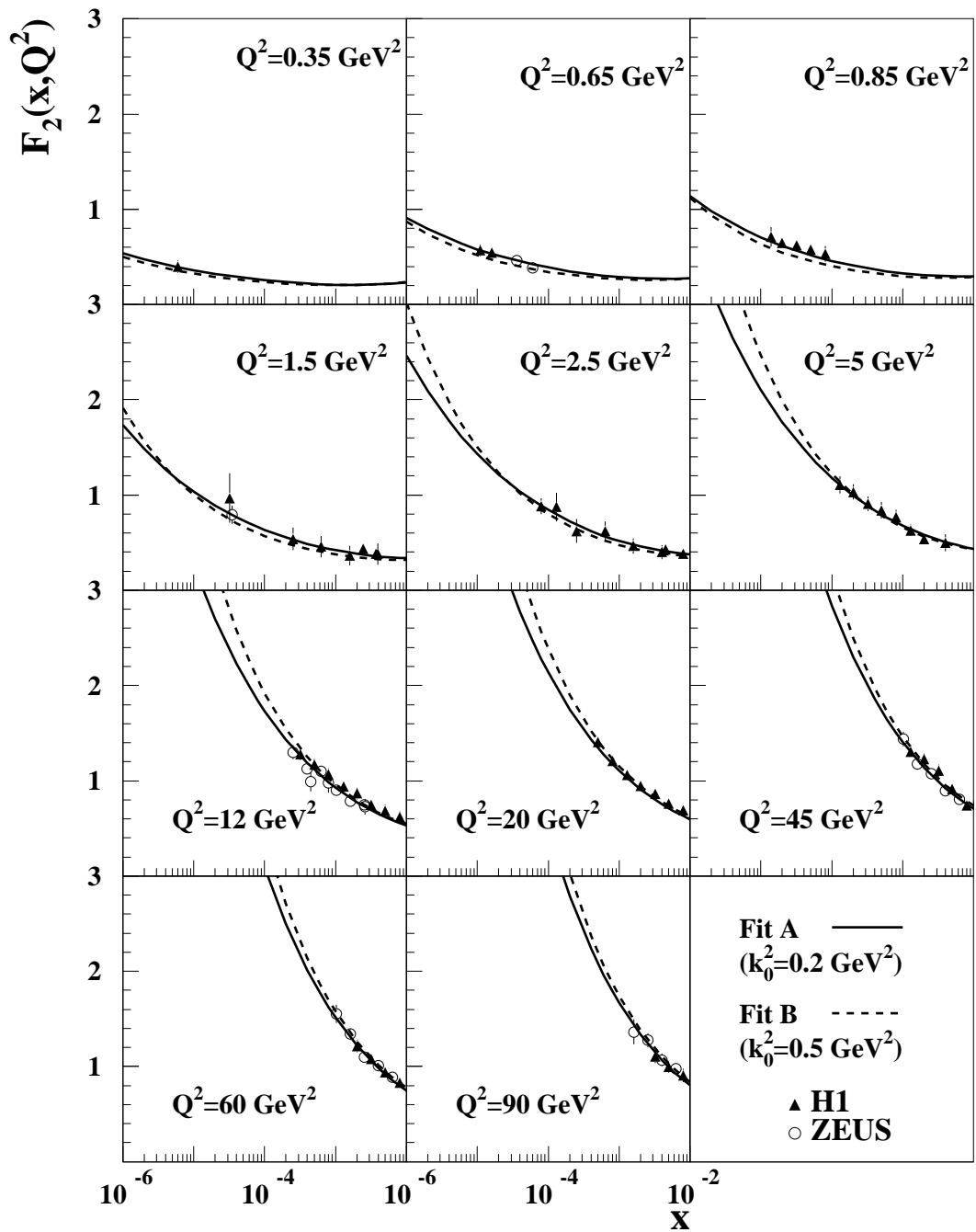
\item[Fig.~4] The description of the $F_2$ data obtained in Fits A and B.  Only a                                             
subset of the data fitted is shown.                                            
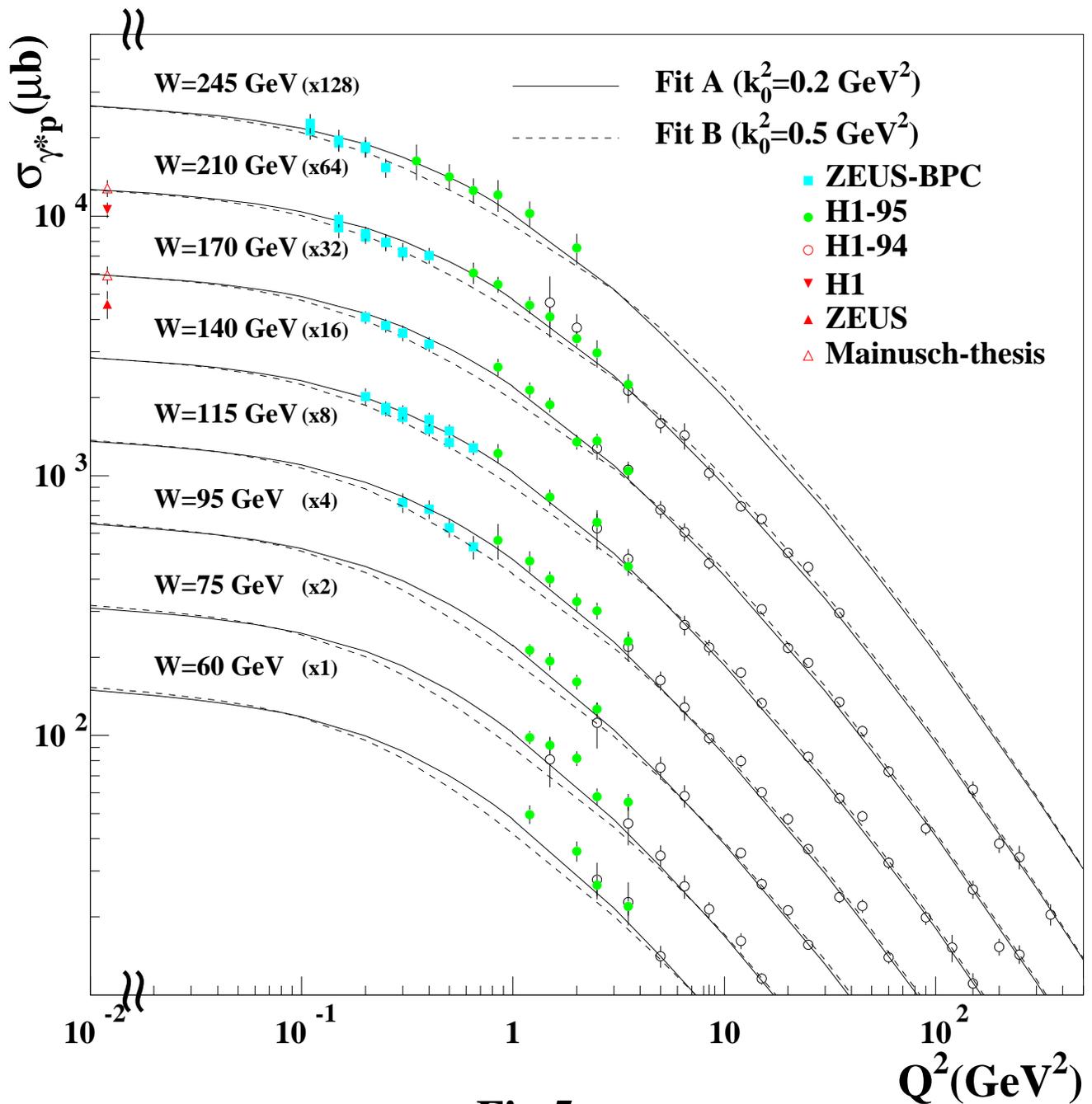
\item[Fig.~5] The curves are the values of the virtual photon-proton cross section                                             
$\sigma_{\gamma^* p}$ of (\ref{eq:a5}) as a function of $Q^2$ for various values of                                             
the energy $W = \sqrt{s}$ corresponding to Fits A and B (multiplied by the factor                                             
shown in brackets).  The data \cite{H1,ZEUS} are assigned to the value of $W$                                             
which is closest to the experimental $W$ bin.  The upper, lower photoproduction                                             
(solid triangular) data points correspond to $W = 210$, 170~GeV and are from the H1                                             
\cite{H1P} and ZEUS \cite{ZP} collaborations respectively.  The open triangular                                      
points are obtained from an analysis of ZEUS photoproduction data reported                                      
in a thesis by Mainusch \cite{SU}.                                      
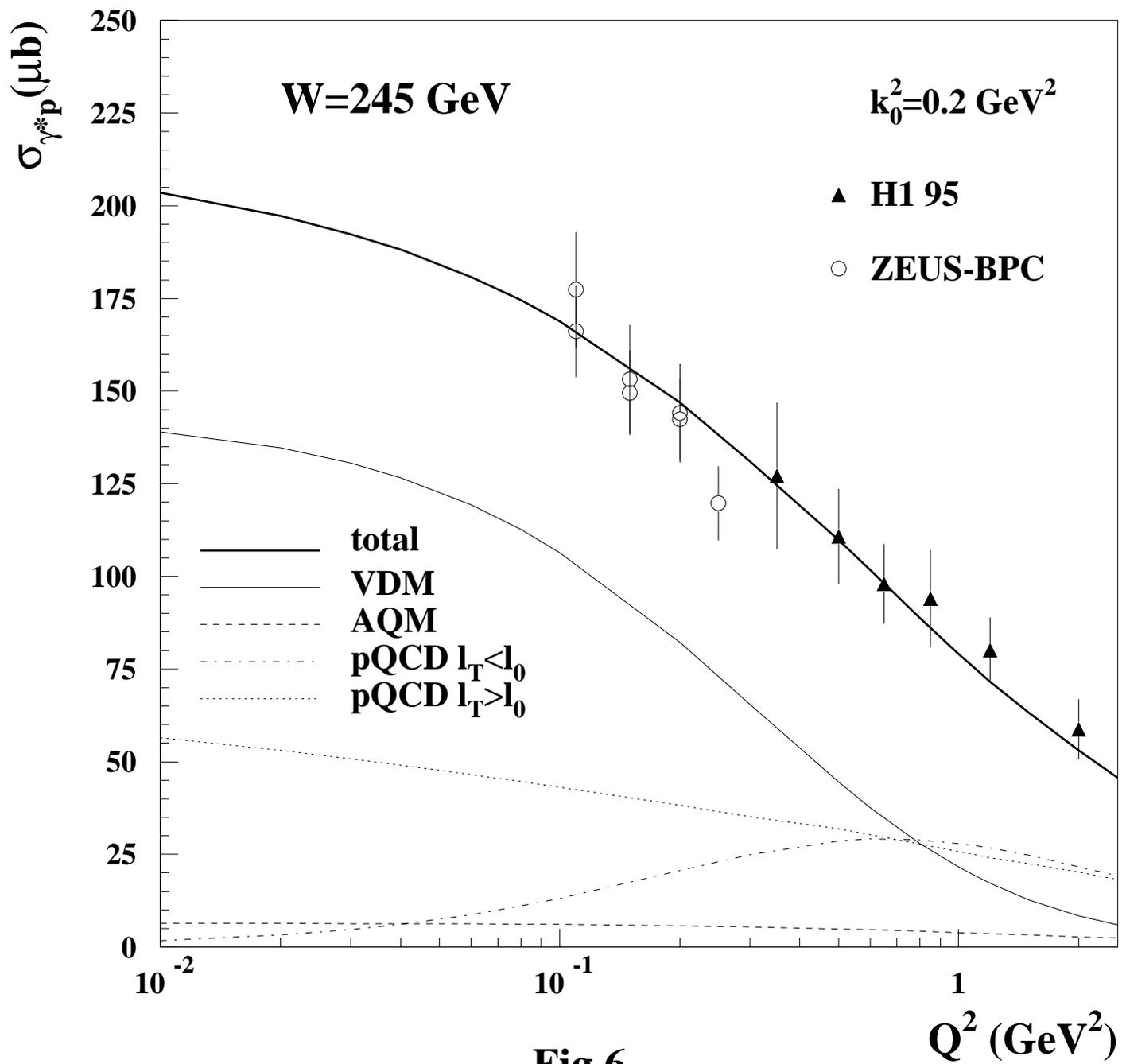
\item[Fig.~6] The various components of $\sigma_{\gamma^* p}$ (as defined in                                             
Section 2.3) shown as a function of $Q^2$ at $W = 245$~GeV for Fit A (with $k_0^2                                             
= 0.2$~GeV$^2$).  The bold curve shows their sum, $\sigma_{\gamma^* p}$,                                             
compared to the HERA measurements.                                            
\item[Fig.~7] The same as Fig.~6 but for Fit B (with $k_0^2 = 0.5$~GeV$^2$).  Th 
e                                             
poorer description of the data in the region $Q^2 \sim 1$~GeV$^2$, as compared to                                             
Fit A, is clearly apparent and can be attributed to the smaller perturbative QCD                                             
component at low gluon $l_T$.                                            
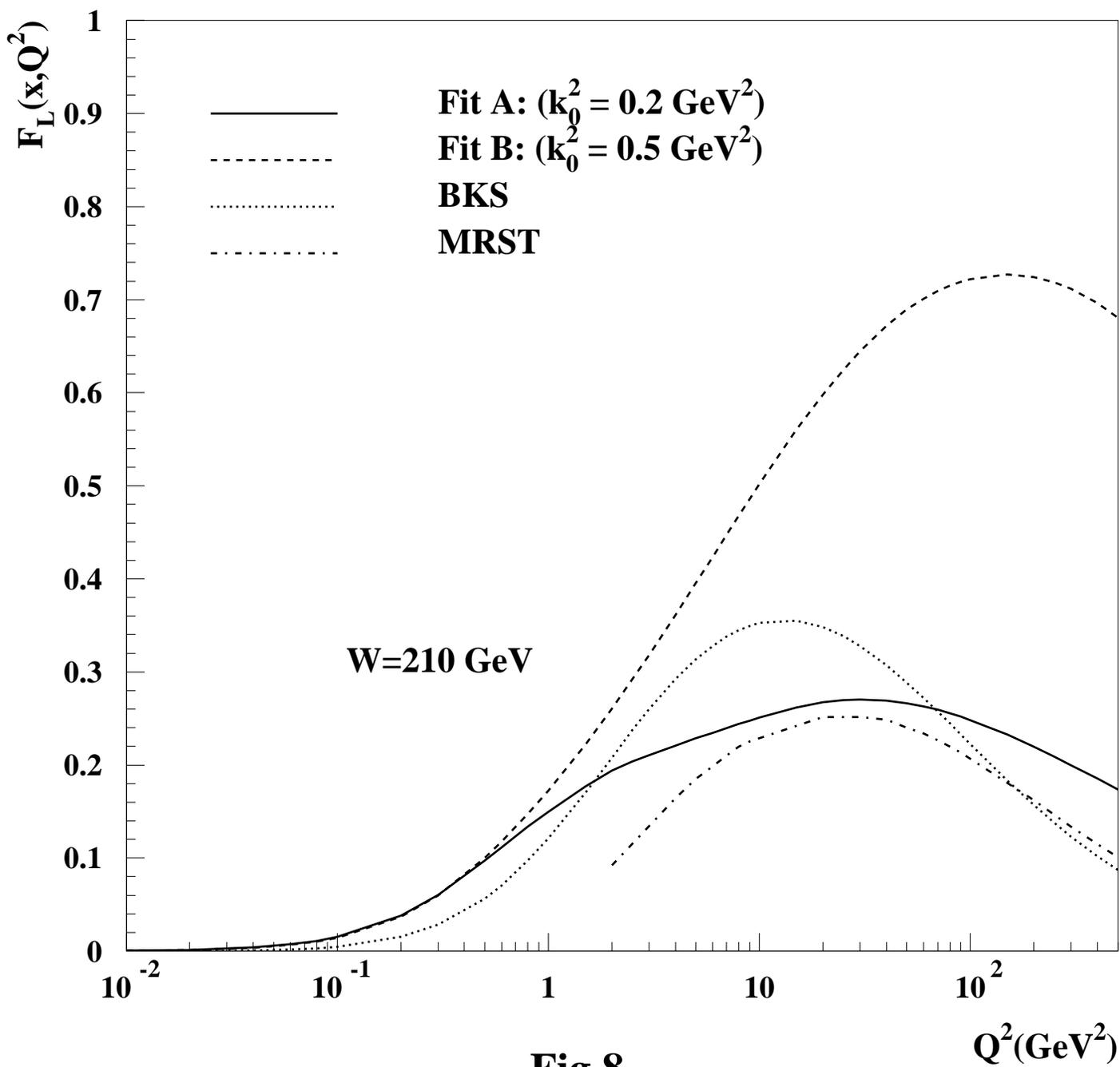
\item[Fig.~8] The predictions for $F_L$ versus $Q^2$ at $W = 210$~GeV from Fits                                             
A, B (with $k_0^2 = 0.2$ and 0.5~GeV$^2$ respectively), together with the values                                             
obtained by Badelek, Kwiecinski and Stasto \cite{BKS} and from the MRST set of                                             
partons \cite{MRST}.                                            
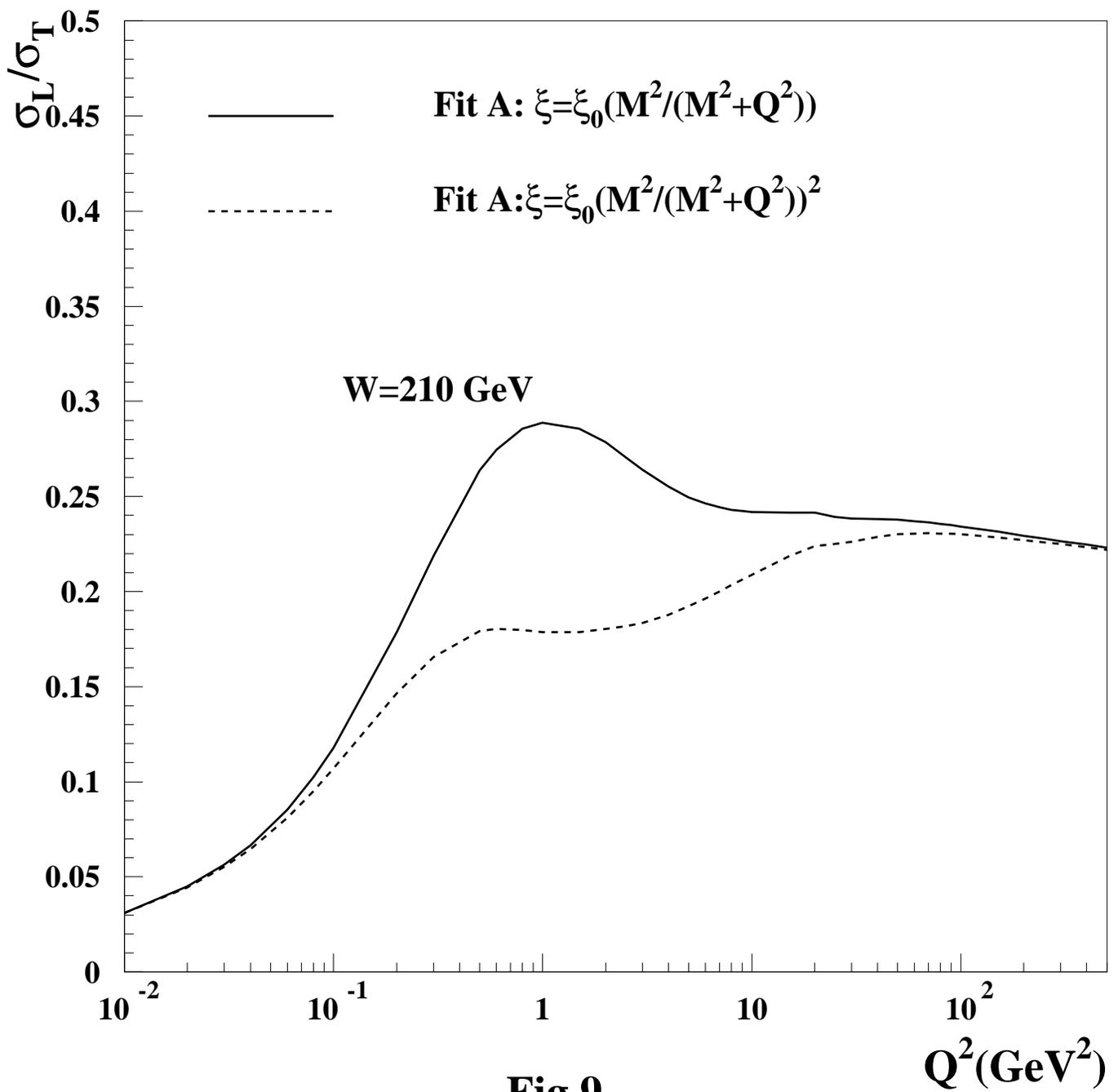
\item[Fig.~9] The dashed curve is the prediction for $\sigma_L/\sigma_T$ versus                              
$Q^2$ at $W = 210$~GeV from Fit A.  For comparison the continuous curve is the                              
prediction obtained using a different choice of the VDM contribution to $F_L$;                              
namely using (\ref{eq:b22}) in the place of (\ref{eq:a22}).                              
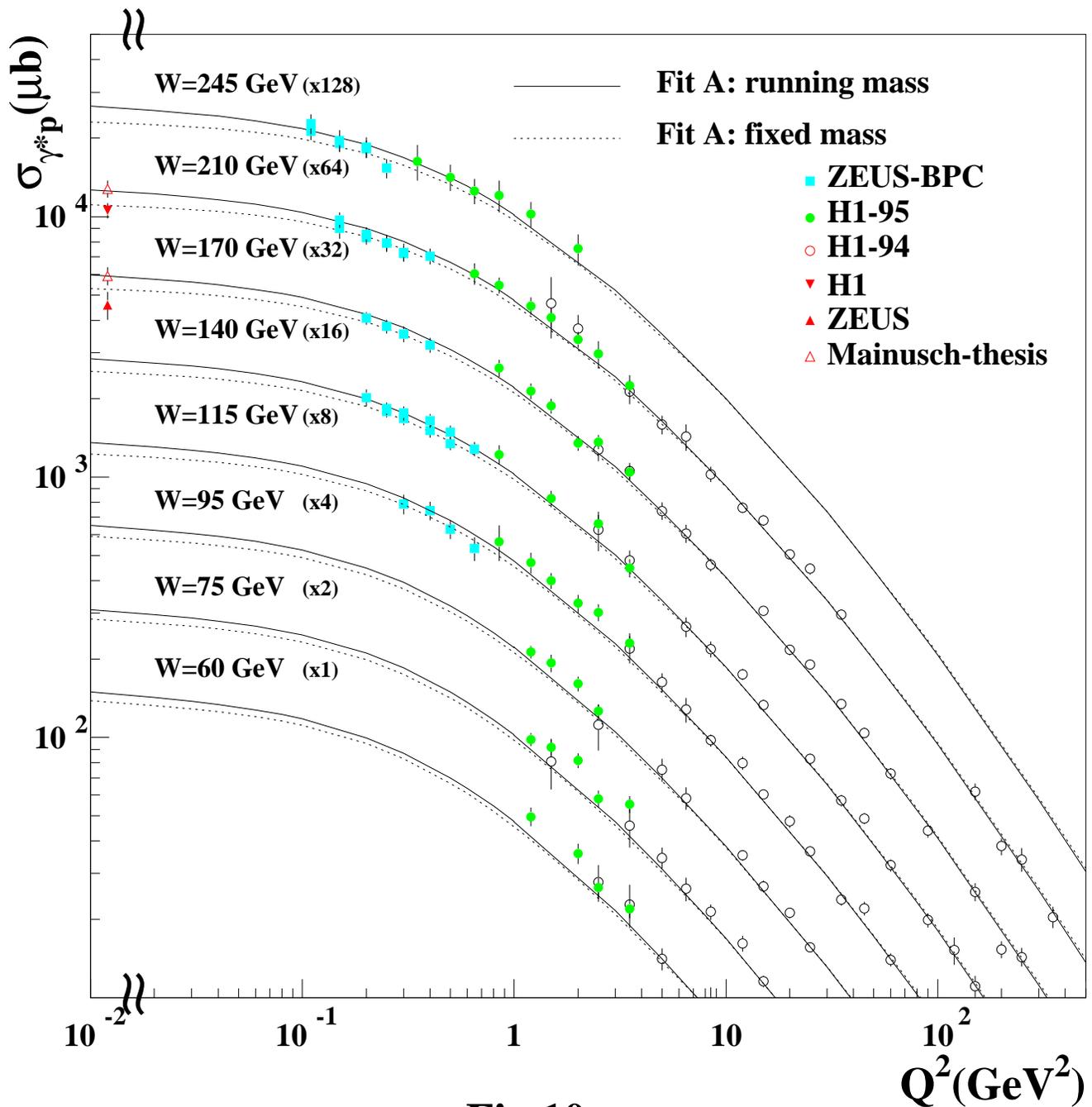
\item[Fig.~10] The dotted curves show the effect of using a (fixed) constituent mass,                                      
$M_0$, in all contributions.  The running mass fit (continuous curves) and the data                                      
are those of Fig.~5.                                      
\end{itemize}                                                
                                                
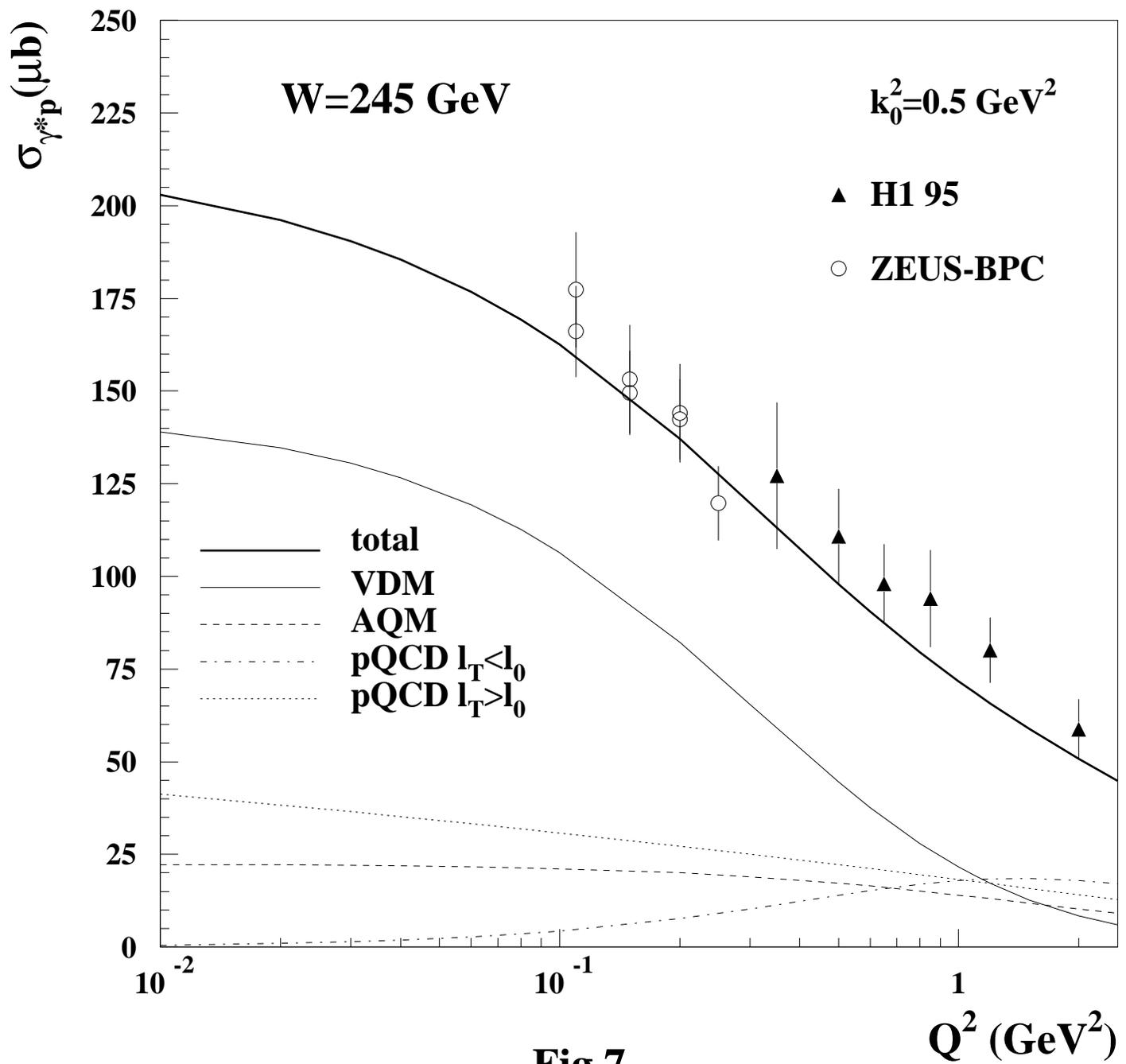
\end{document}